\documentstyle[twocolumn,epsfig,aps]{revtex}
\newcommand{\simle}
{\raisebox{-0.75ex}[-1.5ex]{$\;\stackrel{<}{\sim}\;$}}

\begin{document}

\draft
\title{
Electronic State and Magnetic Susceptibility
in Orbitally Degenerate ($J=5/2$) Periodic Anderson Model}
\author{Hiroshi KONTANI and Kosaku YAMADA$^1$}
\address{
Institute for Solid State Physics, University of Tokyo, 
7-22-1 Roppongi, Minato-ku, Tokyo 106}
\address{$^1$
Department of Physics, Faculty of Science, Kyoto University, 
Kyoto 606-01}

\date{\today}
\maketitle

\def\w{{\omega}}

\begin{abstract}
Magnetic susceptibility in a heavy fermion system is composed of the Pauli 
term ($\chi_{\rm P}$) and the Van-Vleck term ($\chi_{\rm V}$). The latter 
comes from the interband excitation, where $f$-orbital degeneracy is essential.
In this work, we study $\chi_{\rm P}$ and $\chi_{\rm V}$ in the orbitally 
degenerate ($J=5/2$) periodic Anderson model for both the metallic and 
insulating cases. The effect of the correlation between $f$-electrons is 
investigated using the self-consistent second-order perturbation theory.
The main results are as follows. (i) Sixfold degenerate model: both 
$\chi_{\rm P}$ and $\chi_{\rm V}$ are enhanced by a factor of $1/z$
($z$ is renormalization constant). (ii) Nondegenerate model: only 
$\chi_{\rm P}$ is enhanced by $1/z$.
Thus, orbital degeneracy is indispensable for 
the enhancement of $\chi_{\rm V}$.
Moreover, orbital degeneracy reduces the Wilson ratio
and stabilizes the nonmagnetic Fermi liquid state.
\end{abstract}

\vskip 0.6cm
\noindent
KEYWORDS : \
heavy fermion system, Kondo insulator, magnetic susceptibility,
Fermi liquid theory, orbital degeneracy

\vskip 0.6cm

\def\d{\partial}
\def\s{{\sigma}}
\def\e{{\epsilon}}
\def\k{{ {\bf k} }}
\def\p{{ {\bf p} }}
\def\q{{ {\bf q} }}
\def\x{{ {\bf x} }}
\def\w{{\omega}}
\def\l{{\lambda}}
\def\L{{\Lambda}}       
\def\a{{\alpha}}
\def\g{{\gamma}}
\def\G{{\Gamma}}        
\def\v{{\varphi}}
\def\D{{\Delta}}
\def\ddHM{{ \frac1{gM} \frac{\d}{\d H} }}
\def\ddw{{ \frac{\d}{\d \w} }}
\def\dde{{ \frac{\d}{\d \e} }}
\def\i{{ {\rm i} }}
\def\expo{{ {\rm e} }}


In general, the Ce-compound heavy fermion (HF) system
possesses one electron per Ce-atom.
Due to strong $l$-$s$ coupling,
the lowest $f^1$-eigenstate has the angular momentum $J=5/2$.
In real materials, the sixfold states in $J=5/2$
are split into three Kramers doublets (KDs) by a weak crystalline 
electric field (CEF).
The higher KDs strongly affect the electronic structure
because $f$-level splitting is, at most, $\sim O(10^2)$K.
In HF systems, orbital degeneracy (or orbital freedom)
has an essential effect on many physical properties.
For example, it causes the anomalous Hall effect, 
or a quadrapole long range order.
Another interesting feature of the orbitally degenerate systems
is the existence of the Van-Vleck susceptibility $\chi_{\rm V}$,
which is brought about by magnetic interband excitation.
\cite{Zou,Zhang,Anderson,Hanzawa}
$\chi_{\rm V}$ should bring about large residual magnetic susceptibility
in many Kondo insulators 
(CeNiSn, Ce$_3$Bi$_4$Pt$_3$, etc),
 \cite{insulator}
or in superconducting compounds below $T_c$
(UPt$_3$, UBe$_{13}$, etc).
Thus, orbital freedom causes 
various nontrivial phenomena in HF compounds,
and should be studied in earnest.
In addition, orbital freedom also exists in many $d$-electron
systems (V- or Mn-oxide compounds, etc).

In this work, we study the orbitally degenerate ($J=5/2$) periodic Anderson
model ($J=5/2$ PAM), which is the appropriate model for 
Ce-compounds.
In this orbitally degenerate model, the magnetic susceptibility $\chi$ 
is composed of the Pauli term ($\chi_{\rm P}$) and the Van-Vleck term 
($\chi_{\rm V}$).
It is proved that $\chi_{\rm V}$ appears in the case of 
$[{\hat H},{\hat M}]\neq0$, where ${\hat H}$ and ${\hat M}$ are 
the Hamiltonian and the magnetization operators, respectively.
Generally, the analysis of $\chi_{\rm V}$ in the interacting system 
is not easy.
 \cite{Yip,Nakano,Kontani}
In the mean-field approximation (Gutzwiller approximation), 
 \cite{Rice}
where we replace $V_0$ with $\sqrt{z} \cdot V_0$
($z\ll1$ is the renormalization constant),
both $\chi_{\rm P}$ and $\chi_{\rm V}$ are enhanced by $1/z$.
 \cite{Zhang,Hanzawa}
However, this result depends strongly on the assumption that the $f$-electron
level $E_f$ is renormalized by $1/z$ to $E_f^\ast$, 
which is never trivial and should be confirmed in detail.
 \cite{Anderson,Kontani}
For $U>0$, it is also nontrivial 
whether the quasi-particle like spectrum
at $E_f^\ast$ remains coherent or vanishes to become incoherent.

The aim of this work is to study the electronic structure and the magnetic
susceptibility of $J=5/2$ PAM beyond the mean-field theory.
Here, we perform the numerical calculation by using the 
self-consistent second-order perturbation theory (SC-SOPT) 
with respect to (w.r.t.) $U$.
In principle, (SC-)SOPT is valid only for the weak-coupling regime.
Nonetheless, it is known to be very useful for 
discussing strong coupling properties of, e.g., the
impurity Anderson model \cite{Yamada2} and the Hubbard model.
 \cite{Schweitzer}
In addition, we use the local approximation ($d=\infty$ approximation),
\cite{Metzner}
which is known to work well for nondegenerate PAM.
\cite{Schweitzer,Mutou}
We check the higher correction terms w.r.t. $1/d$ for $J=5/2$ PAM,
and find that the local approximation works better due to
the orbital degeneracy.
\cite{explain2}

We find that, for the six-fold degenerate case (no CEF case),
both $\chi_{\rm P}$ and $\chi_{\rm V}$
are enhanced by the enhancement factor at the Fermi energy
within SC-SOPT.
This result coincides with that of our analytical study
on the basis of the Fermi liquid theory for the strong 
correlation region, given in ref. \cite{Kontani}.
This numerical work is complementary to the analytical work in
ref. \cite{Kontani}.
We also confirm the renormalization of $E_f$ 
towards the Fermi energy, $\mu$.
This result is consistent with the enhancement of $\chi_{\rm V}$.

\begin{figure}
\epsfxsize=50mm \epsffile{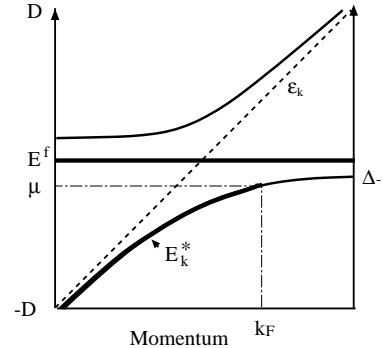}
\caption
{Schematic electronic structure of $J=5/2$ PAM for $U=0$
in the metallic state ($\mu<\Delta_{-}$).
Fourfold degeneracy remains at $E_f$.
Band insulating state is realized for $\Delta_{-}<\mu<E_f$.
}
\label{fig:band}
\end{figure}

The Hamiltonian for $J=5/2$ PAM is given as
\begin{eqnarray}
& & H = \sum_{\k \sigma} \epsilon_{\k} c_{\k \sigma}^\dagger c_{\k \sigma}
       + \sum_{\k M} E_f f_{\k M}^\dagger f_{\k M}
       + \sum_{M\k \sigma} ( V_{\k M\sigma}^\ast f_{\k M}^\dagger c_{\k \sigma}
 \nonumber \\
& &\ \ \ \          + {\rm h.c.} )  
 + \frac U2 \sum_{\k \k ' \q M\neq M'}
 f_{\k - \q M}^\dagger f_{\k '+ \q M'}^\dagger f_{\k 'M'} f_{\k M} ,
 \label{eqn:hamiltonian}
\end{eqnarray}
where $c_{\k \sigma}^\dagger$ ($f_{\k M}^\dagger$)
is the creation operator of a conduction electron ($f$-electron),
and $\e_\k$ is the $\k$-dependent spectrum for conduction electrons.
$M$ is the eigenvalue of $J_z$ ($M=5/2,3/2,\cdots,-5/2$).
The $c$-$f$ mixing potential $V_{\k M\s}$ in (\ref{eqn:hamiltonian}) 
is given by
$V_{\k M\s}= V_0 \cdot \sqrt{4\pi/3} \sum_m -\s \sqrt{ (7/2-M \s)/7 }
 \delta_{m,M-\s/2} Y_{l=3}^m (\theta_k ,\varphi_k )$,
where
$Y_{l=3}^m(\theta_k ,\varphi_k )$ is the spherical harmonic function.
We impose the Zeeman energies for only $f$-electrons,
$E_{M}^f =E_f + g \mu_B M \cdot H$,
where $g$ is Lande's $g$-factor ( $g=6/7$ for $J=5/2$ ),
and we set $\mu_B=1$ hereafter.
The electronic structure in the case of $U=0$ is shown schematically
in Fig. \ref{fig:band}.
$\Delta_{-}=E_{k=2\pi}$, and
$E_k$ is the quasi-particle spectrum, which is the bonding orbit
brought about by $\e_\k$ and $E_f$.
There remain fourfold nonbonding orbits at $E_f$.

\begin{figure}
\epsfxsize=60mm \epsffile{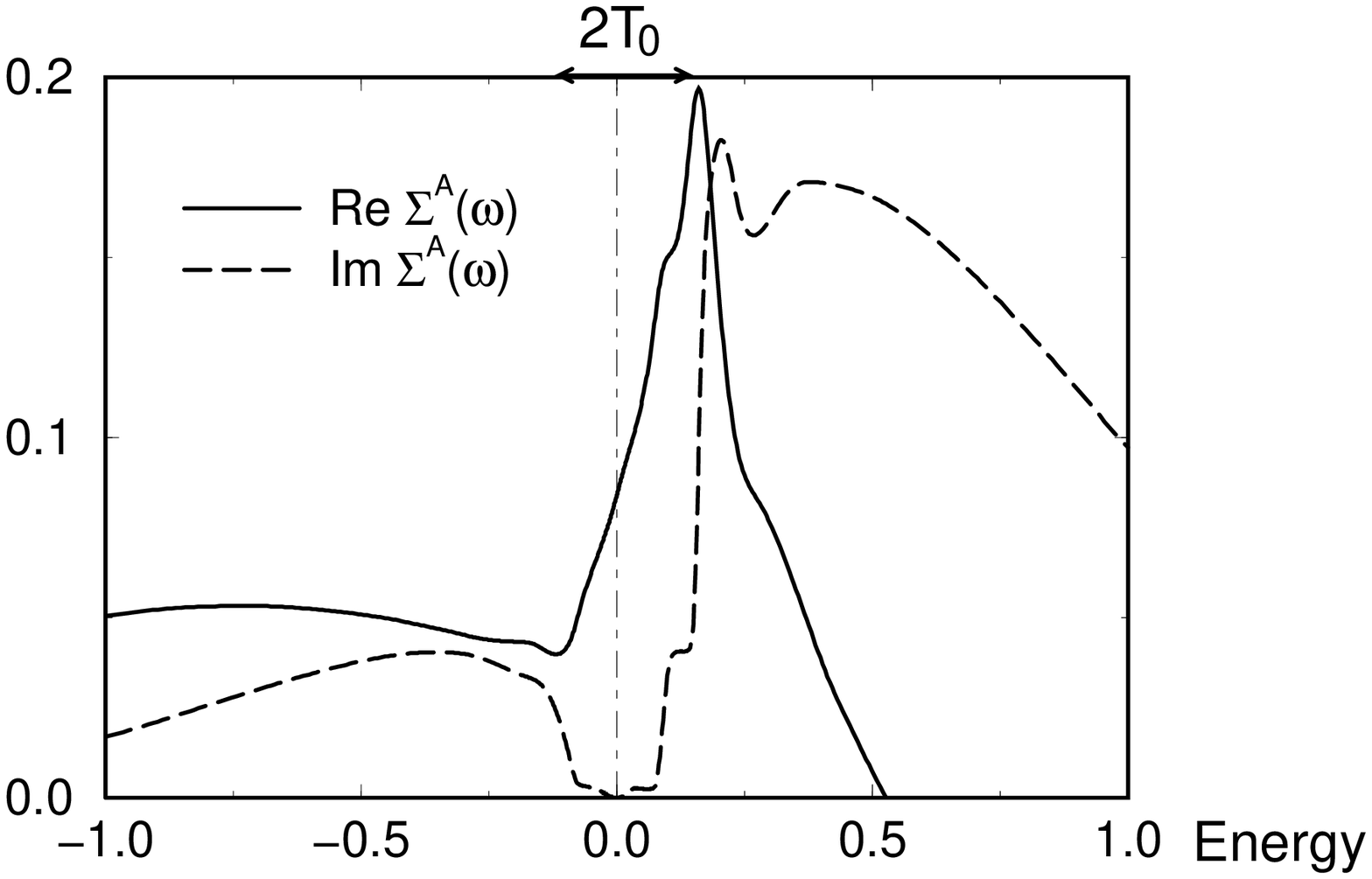}
\caption
{The self-energy of the metallic $J=5/2$ PAM by SC-SOPT 
for $U=0.3$, where $1/z(0)=1.53$.
For $|\w|<0.02$, the relation
${\rm Im}\Sigma^A(\w) = U^2\cdot (5\pi/2) \rho(0)^3 \cdot \w^2
\approx 2.4\cdot\w^2$ is satisfied.
Other parameters  are
$\e_\k= -1+2\cdot(\k/\pi)^2$, $V_0=0.4$, 
$E_f=-0.25$ and $\mu=-0.38$.
In each stage of self-consistent calculation, we choose $E_f$
so that $\mu$ is unchanged.
Note that $E_f^\ast \approx 0.1 < T_0$ (see Fig. 3).
}
\label{fig:selfenergy}
\end{figure}

The local Green's function $g(\w)$,
with which the self-energy $\Sigma(\w)$ is constructed in the local 
approximation, is given at $H=0$ by
 \cite{Kontani}
\begin{eqnarray}
& &g(\w)= \frac13 G(\w) + \frac23 G^{f}(\w), 
 \label{eqn:green1} \\
& &G(\w)= \frac 1N \sum_k \left({1/G^{f}(\w)- 3V^2/(\w+\mu-\e_\k)}
 \right)^{-1},  \label{eqn:green2} \\
& &G^{f}(\w)= \left({\w+\mu-E_f-\Sigma(\w)} \right)^{-1}
 \label{eqn:green3}.
\end{eqnarray}
and the $\w$-dependent renormalization factor in the Fermi liquid theory is
\begin{eqnarray}
 z(\w)= \left(1- \d \Sigma_M(\w) /\d\w \right)^{-1},
\end{eqnarray}
This $\w$-dependence is neglected in the mean-field treatment.
Note that both $g(\w)$ and $\Sigma(\w)$ are diagonal and 
independent of $M$ in the case of no CEF.
 \cite{Kontani}
Considering the Pauli principle,
the self-energy by SC-SOPT in the local approximation is given by
\begin{eqnarray}
\Sigma(\w)= \frac{5U}{6}n^f - 5U^2 \int \!\! \int 
 \frac{{\rm d}\e {\rm d}\e'}{4\pi^2}g(\e)g(\e')g(\w+\e-\e')
 \label{eqn:self}
\end{eqnarray}
at $T=0$.
Figure \ref{fig:selfenergy} shows the calculated self-energy $\Sigma(\w)$.
The $\w$-dependence of $\Sigma(\w)$ is complicated,
and the energy region, where renormalization takes place
by a factor of $1/z(\w)> 1$, is limited to within $2T_0$ from the Fermi energy.
(We call $T_0$ the coherent energy in the system.)
These features are not derived by the mean-field treatment.

\begin{figure}
\epsfxsize=60mm \epsffile{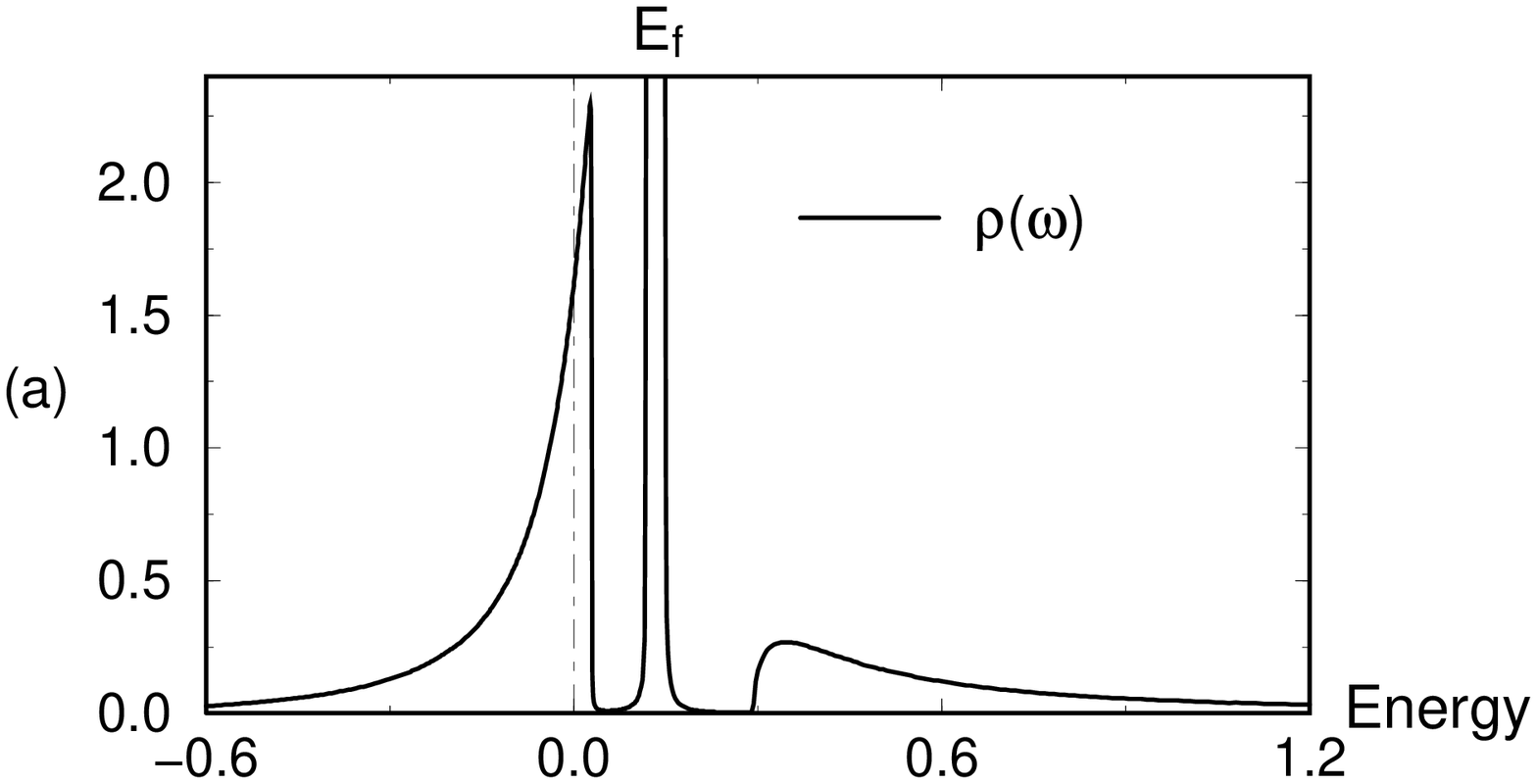}
\epsfxsize=60mm \epsffile{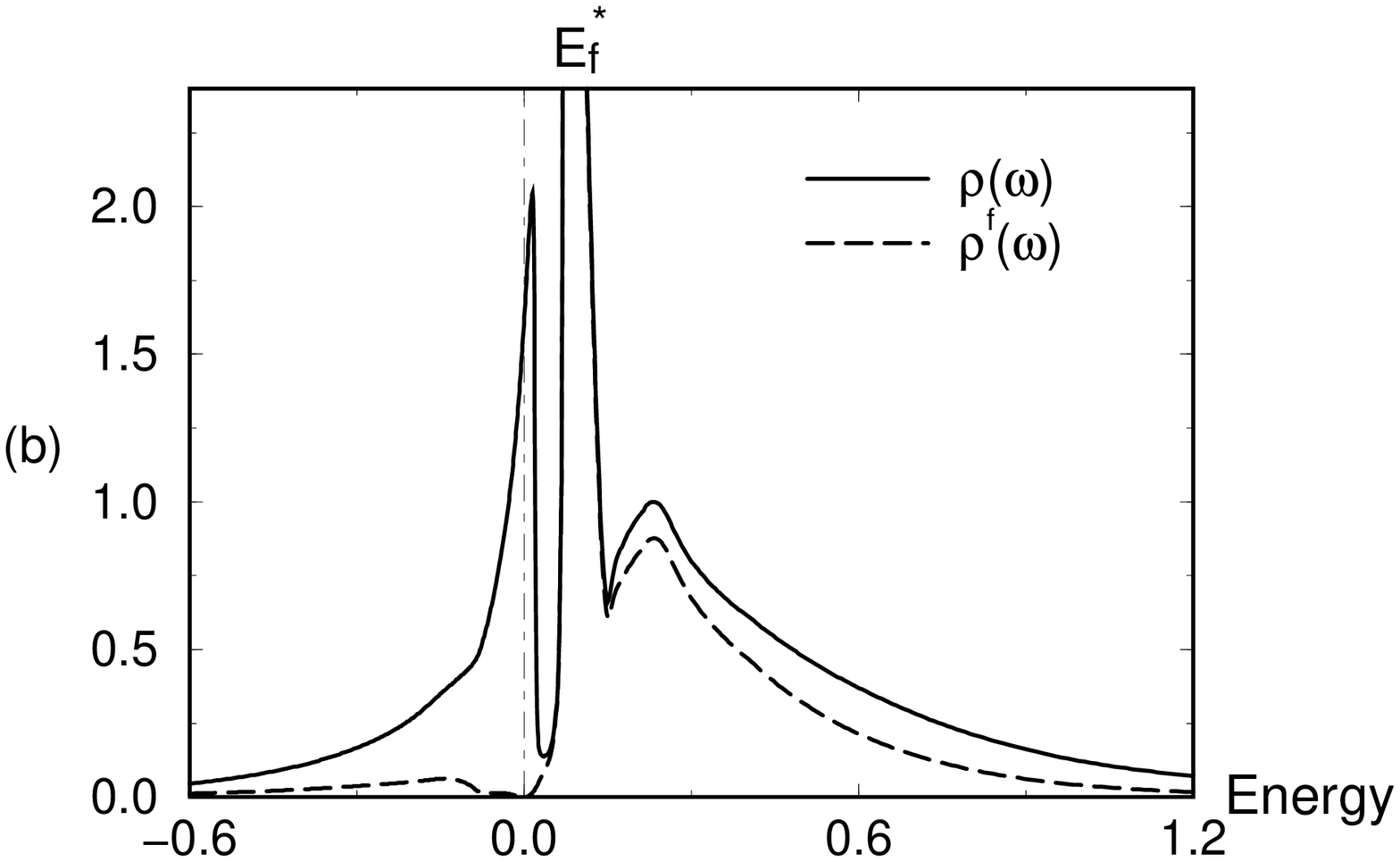}
\caption
{The $f$-electron DOS of the metallic $J=5/2$ PAM obtained by 
SC-SOPT for (a) $U=0$ and (b) $U=0.3$, where $1/z(0)=1.53$.
Other parameters are the same as 
in Fig.2.
We can see that $\rho^f(\w)$ for $U=0.3$ is 
spread into a wider energy range.
$\rho(0)$ is unchanged by $U$ in the local approximation.
}
\label{fig:dos}
\end{figure}

Figure \ref{fig:dos} shows the $f$-electron density-of-states (DOS),
$\rho(\w)= -\frac 1\pi {\rm Im}\ g(\w+\i\delta)$,
obtained by using (\ref{eqn:green1})-(\ref{eqn:green3}) and (\ref{eqn:self}).
We also introduce the DOS for the nonbonding orbits,
$\rho^f(\w)= -\frac 1\pi \frac23 {\rm Im}\ G^{f}(\w+\i\delta)$.
In the numerical calculation,
we imposed weak $k$-dependence on $E_f$ for convenience.
The noteworthy points are as follows.
(i) $E_f$ is renormalized toward $\mu$ to be $E_f^\ast$,
and its weight is reduced.
We find that the relation $|E_f^\ast-\mu| \approx z(0)\cdot |E_f-\mu|$
holds well for smaller $U$ where SC-SOPT is adequate.
(ii) Broad satellites (or shoulders) are formed on both sides of $\mu$,
which are attributed to Mott-Hubbard excitation.
Large part of $\rho^f(\w)$ is transferred to the satellites, 
which is localized at $E_f$ in the case of $U=0$.
This feature is not derived by the mean-field theory.
The renormalization of $E_f$ observed in our calculation
means that $E_f^\ast$ lies inside the 
coherent region, and $\chi_V$ is expected to be enhanced.

Next, we calculate the uniform magnetic susceptibility for the
following two typical cases. \\
{\bf (i) Case 1} (there is no CEF and three KDs are degenerate):
In the previous work based on the Fermi liquid theory,
 \cite{Kontani} 
we found the following result in the strong coupling region:
\begin{eqnarray}
& &\chi/\chi^0 = 1/z_H(0)
 \ \ \ \  \ \ \ \ {\rm for \ the \ metallic \ case}, 
 \label{eqn:chi-m} \\
& &\chi/\chi^0 = 1/z_H(\Delta_{-}^\ast)
 \ \ \ \ \ {\rm for \ the \ insulating \ case}, 
 \label{eqn:chi-i} \\
& & \ \ \ 1/z_H(\w) \equiv 1+ \ddHM \Sigma_{M}(\w),
\end{eqnarray}
whose deviations are of order $\sim O(z^0)$, and 
$1/z_H(\w)= 1+ \frac1M \frac{\d}{\d H}\Sigma_M(\w)$ 
is the magnetic enhancement factor in the Fermi liquid theory.
In general, $1/z(0) \simle 1/z_H(0)$
through the Landau parameter $F_0^a<0$.
 \cite{Kontani}
Thus, $\chi_{\rm insulator}=\chi_V$ is enhanced.
In the metallic state, total susceptibility $\chi$
is given by the $k$-limit of the dynamical susceptibility $\chi_k(\w)$
(i.e., $\chi= \lim_{k\rightarrow0}\chi_\k(0)$)
and $\chi_{\rm V}$ should be defined by the $\w$-limit of $\chi_k(\w)$.
In the insulating state, the two limits coincide.

\begin{figure}
\epsfxsize=60mm \epsffile{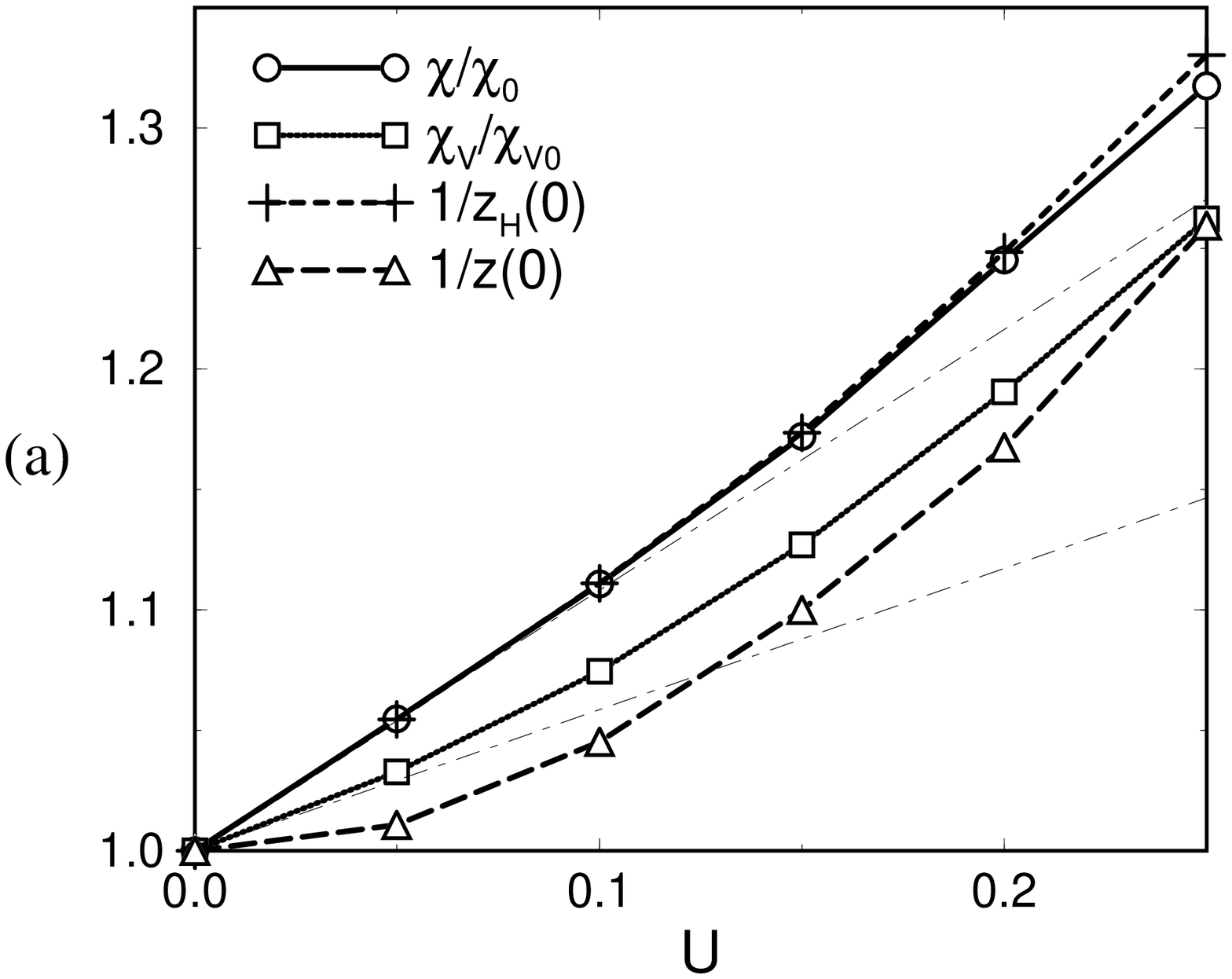}
\epsfxsize=60mm \epsffile{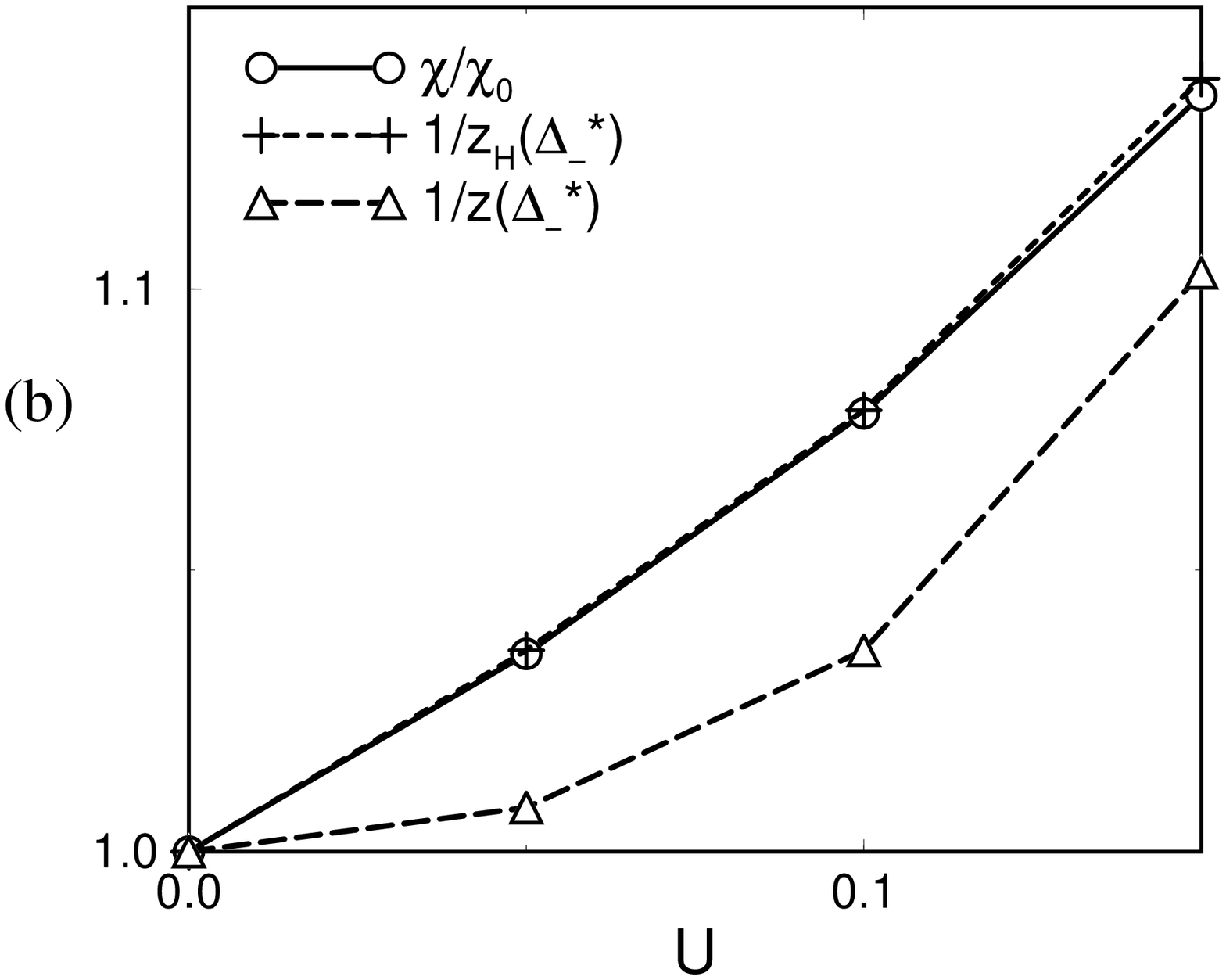}
\caption
{In case 1 (sixfold degenerate case):
$\chi$, $1/z_H$ and $1/z$ obtained using SC-SOPT for 
(a) the metallic case and for (b) the insulating case.
For (a), we used the same parameters as those
in Fig.2.
For (b), parameters are
$\e_\k= -1+2\cdot(\k/\pi)^3$, $V_0=0.4$ and $E_f=0$.
We also imposed $\mu= (\Delta_{-}^\ast+E_f^\ast)/2$
which should be satisfied in the case of a pure insulator at $T=0$.
}
\label{fig:susc-j}
\end{figure}

Using the above relations, we calculate $\chi$ 
in a manner consistent with the self-energy obtained by SC-SOPT,
so as to satisfy the conservation law (i.e., the Ward identity) proposed in
ref. \cite{Baym}.
Furthermore, we compare it with $1/z_H(0)$ (or $1/z_H(\Delta_{-}^\ast)$)
derived using SC-SOPT.
Here, $\ddHM g_M(\w)|_{H=0}$ is given by (5.12) in 
ref. \cite{Kontani}, and is complicated because
${\hat M}$ has interband components.
(Note that $[{\hat H},{\hat M}]\neq0$.)
The results for (a) the metallic case and (b) the insulating case
at $T=0$ are shown in Fig. \ref{fig:susc-j}.
We can see that 
relations (\ref{eqn:chi-m}) and (\ref{eqn:chi-i}) hold with high accuracy
for small $U$, where the result obtained using SC-SOPT is reliable.
We consider the deviation from (\ref{eqn:chi-m}) and (\ref{eqn:chi-i})
for large $U$ in Fig. \ref{fig:susc-j} to be artificial because
it is of the order, at least, $O(U^3)$.
Taking account of the analitical result in the strong coupling region,
 \cite{Kontani}
we conclude that the relation 
(\ref{eqn:chi-m}) and (\ref{eqn:chi-i}) hold irrespective of 
the strength of the correlation.
Moreover, Fig. \ref{fig:susc-j}(a) clearly shows the enhancement of 
$\chi_{\rm V}$ in the metallic state.

\begin{figure}
\epsfxsize=60mm \epsffile{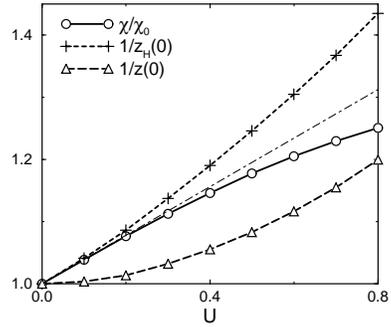}
\caption
{In case 2 (twofold degenerate case):
$\chi$, $1/z_H(0)$ and $1/z(0)$ obtained by SC-SOPT for the insulating case.
Here we used a symmetric PAM
($\e_\k= -1+2\cdot(\k/\pi)^3$, $V_0=0.4$, 
 $E_f=\mu=0$).
}
\label{fig:susc-s}
\end{figure}

{\bf (ii) Case 2} (where only the lowest KD is active due to strong CEF):
The system reduces to the familiar nondegenerate PAM.
In this model, $\chi_{\rm V}$ is finite unless the $g$-values for
$f$-electrons ($g_f$) and for $c$-electrons ($g_c$) are the same.
 \cite{Nakano}
Here, we set $g_c=0$. The result of the  mean-field approximation is
\begin{eqnarray}
& &\chi_{\rm P} = g_f^2 \cdot \rho(0)/z, 
 \label{eqn:chi2-p} \\
& &\chi_{\rm V} = g_f^2 \cdot \rho_c \cdot
{\rm cot}^{-1}(\sqrt{z}V/(\mu-\e_{k_F})),
 \label{eqn:chi2-v}
\end{eqnarray}
where $\rho(\w)$ ($\rho_c$) is the $f$($c$)-electron DOS.
($\chi_{\rm P}^{U=0} \gg \chi_{\rm V}^{U=0}$ 
because $\rho(0)=V^2/(E_f-\mu)^2\cdot \rho_c$.)
Thus, the enhancement of $\chi_{\rm V}$ is of the order $O(1)$
even in the mean-field approximation.
This behavior is recognized more accurately in quantum Monte Carlo 
simulation within the $d=\infty$ approximation.
 \cite{Mutou2,Saso}
Here, we calculate $\chi$ and $1/z_H$ for the insulating state using SC-SOPT,
and show the result in Fig. \ref{fig:susc-s}.
We can see that $\chi$ deviates from $1/z_H$ in the $U^2$-term.
Especially, the sign of curvature of $\chi$ w.r.t. $U$ 
and that of $1/z_H$ are opposite.

\begin{figure}
\epsfxsize=50mm \epsffile{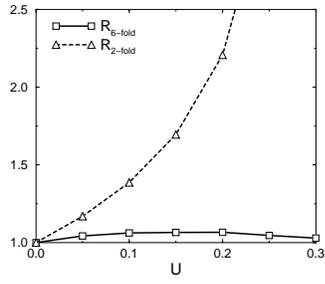}
\caption
{$U$-dependence of the Wilson ratio obtained using SC-SOPT for case 1
($R_{\rm 6\mbox{-}fold}$) and for case 2  
($R_{\rm 2\mbox{-}fold}$).
In both cases, we used the same parameters (and $\rho(0)$) as those 
in Fig 2.
}
\label{fig:wilson}
\end{figure}

Here, we compare several properties between case 1 and case 2.
The sign of the curvature of $\chi_{\rm V}$ w.r.t. $U$
are opposite in the two case.
Thus, interestingly, orbital degeneracy is indispensable for
the large enhancement of $\chi_{\rm V}$.
Next, we discuss the coherent energy $T_0$ for both cases
(see Fig.\ref{fig:dos}).
We can see that $T_0$ in case 1 (sixfold degenerate)
is much larger than that in case 2 (twofold degenerate)
if $1/z$ is the same in both cases.
Roughly speaking, within SC-SOPT, we can see that
$T_0/z \sim |E_f-\mu|$ for case 1 and $T_0/z \sim |\Delta_{-}-\mu|$
for case 2 in the metallic system, where $E_f$ and $\Delta_{-}$ 
are not renormalized values.
This result indicates that the paramagnetic state is more stable in case 1.
Finally, we discuss the Wilson ratio $R$ for both cases,
where $R=(\chi/\chi^{U=0})\cdot z$ in the local approximation.
In Fig. \ref{fig:wilson}, we plot the $U$-dependence of $R$ 
derived using SC-SOPT for both case 1 and case 2.
In case 1 , $R$ increases much more slowly w.r.t. $U$.
As a result, the orbital degeneracy stabilizes the paramagnetic
state against (ferro-)magnetic instability.
This tendency is consistent with the discussion on $T_0$ above,
and it is also found within the Gutzwiller approximation.
 \cite{Rice}

In this work, we have studied the $f$-electron DOS and $\chi$ in $J=5/2$ PAM
by applying the SC-SOPT, in order to study many-body effects
beyond the mean-field approximation.
We showed that the Coulomb interaction drastically changes the 
whole electronic structure.
This change causes the enhancement of $\chi_{\rm V}$, which is the
interband contribution away from the Fermi level.
In Fig. \ref{fig:susc-j},
we found that the relations (\ref{eqn:chi-m}) and (\ref{eqn:chi-i})
hold with high accuracy, which means both $\chi_{\rm P}$ and $\chi_{\rm V}$
are enhanced by the magnetic enhancement factor, $1/z_H$.
This result coincides with our 1996 analytical study based on the 
Fermi liquid theory.
Interestingly, the enhancement of $\chi_{\rm V}$ is closely related to
the orbital degeneracy because it is not observed in nondegenerate PAM
(see Fig. \ref{fig:susc-s}).
Moreover, comparing the Wilson ratio between 
$J=5/2$ PAM and nondegenerate PAM, we can see within the scope of 
SC-SOPT that orbital degeneracy stabilizes the nonmagnetic state.

We would like to express our thanks to Kazuo Ueda for his valuable comments.
We also thank Tetsuro Saso and Tetsuya Mutou for their useful discussions.
This work is financially supported by a Grant-in-Aid for Scientific
Research on Priority Areas from the Ministry of Education, 
Science, Sports and Culture.



\end{document}